\documentclass[a4paper,twoside,english,5p,10pt,sort&compress]{elsarticle}
\usepackage[LGR,T1]{fontenc}
\usepackage{graphicx}
\usepackage{esint}

\makeatletter

\pdfpageheight\paperheight
\pdfpagewidth\paperwidth

\DeclareRobustCommand{\greektext}{%
  \fontencoding{LGR}\selectfont\def\encodingdefault{LGR}}
\DeclareRobustCommand{\textgreek}[1]{\leavevmode{\greektext #1}}
\ProvideTextCommand{\~}{LGR}[1]{\char126#1}

\providecommand{\tabularnewline}{\\}

\makeatother

\usepackage{babel}
\begin{document}

\begin{frontmatter}{}

\title{Reactor antineutrino shoulder explained by energy scale nonlinearities?}

\author{G.~Mention$^{(a)}$, M.~Vivier$^{(a)}$, J.~Gaffiot$^{(a)}$,
T.~Lasserre$^{(a,b)}$, A.~Letourneau$^{(a)}$, T.~Materna$^{(a)}$ }

\address{$^{(a)}$IRFU, CEA, Université Paris-Saclay, F-91191 Gif-sur-Yvette,
France}

\address{$^{(b)}$AstroParticule et Cosmologie, Université Paris Diderot,
CNRS/IN2P3, CEA/DRF/IRFU, Observatoire de Paris, Sorbonne Paris Cité,
75205 Paris Cedex 13, France.}
\begin{abstract}
The Daya Bay, Double Chooz and RENO experiments recently observed
a significant distortion in their detected reactor antineutrino spectra,
being at odds with the current predictions. Although such a result
suggests to revisit the current reactor antineutrino spectra modeling,
an alternative scenario, which could potentially explain this anomaly,
is explored in this letter. Using an appropriate statistical method,
a study of the Daya Bay experiment energy scale is performed. While
still being in agreement with the $\gamma$ calibration data and $\phantom{}^{12}{\rm B}$
measured spectrum, it is shown that a ${\cal O}(1\%)$ deviation of
the energy scale reproduces the distortion observed in the Daya Bay
spectrum, remaining within the quoted calibration uncertainties. Potential
origins of such a deviation, which challenge the energy calibration of these
detectors, are finally discussed.
\end{abstract}

\end{frontmatter}{}

\section{Introduction}

Reactor antineutrino experiments have played a leading role in neutrino
physics starting with the discovery of the electron antineutrino in
1956~\cite{RC53,RC59}, through the first observed oscillation pattern
in KamLAND~\cite{KL}, up to recent high precision measurements on
the $\theta_{13}$ mixing angle~\cite{DC,DB,RN}. Future projects
JUNO~\cite{JN} and RENO50~\cite{RN50} even aim at reaching sub-percent
accuracy on $\theta_{12}$ on top of solving the neutrino mass hierarchy
puzzle. However, two anomalies in the measured antineutrino spectra
are being observed. The first is an overall rate deficit around 6\%
known as ``The reactor antineutrino anomaly''~\cite{RAA}. The
second one is a shape distortion in the 4\textendash 6~MeV region,
often quoted as a ``bump'' or ``shoulder'' in the spectra. It
should be particularly stressed out that the relation between these
two anomalies is not straightforward since shape distortion does not
necessarily imply a change in the total rate.

This letter focuses on the second anomaly. In Section~\ref{sec:Reactor-spectra-comparison},
a quantitative comparison of four reactor antineutrino experiments
(Bugey~3~\cite{B3}, Daya Bay~\cite{DB}, Double Chooz~\cite{DC}
and RENO~\cite{RN}) is performed to demonstrate their incompatibility,
thus questioning nuclear effects as a common origin, as proposed in~\cite{HA}.
The next sections are dedicated to the study of an alternative scenario
accounting for the observed distortion. Section~3 reviews the energy
scale determination in such reactor antineutrino experiments. Section~4
introduces a combined analysis of the Daya Bay calibration and reactor
antineutrino data. Results are presented in section~5 and show that
a~1\% unaccounted break at 4~MeV in the energy scale can reproduce
the observed antineutrino spectrum and still comply with calibration
data within uncertainties. Section~6 discusses possible origins of
such an energy nonlinearity and especially questions calibration of
such detectors.

\section{Reactor spectra comparison\label{sec:Reactor-spectra-comparison}}

\subsection{On statistical compatibility of reactor spectra}

\begin{figure}
\begin{centering}
\includegraphics[scale=0.48]{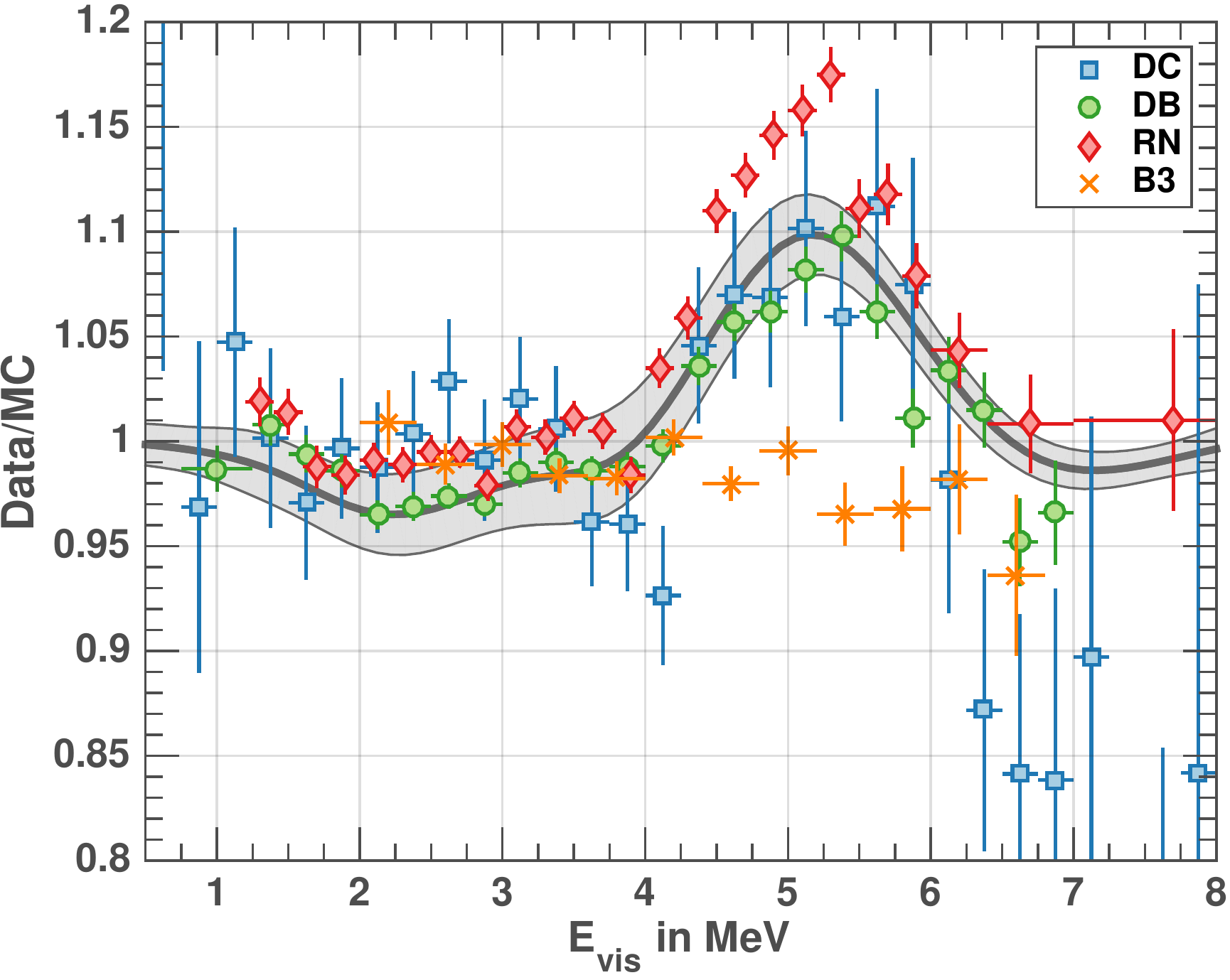}\hspace{1ex}
\par\end{centering}
\caption{\label{fig:Ratios}Ratio of observed reactor antineutrino spectra
to current best predictions~\cite{HU,MU}. Despite similar fuel compositions,
Double Chooz (DC), Daya Bay (DB) and RENO (RN) display significant
deviations around 5~MeV, while Bugey~3 (B3) does not. The global
best fit is indicated by the gray curve. A dedicated Monte Carlo simulation
shows the compatibility p-value is below $10^{-5}$ at 99\% CL.\@
The uncertainty on the best fit is illustrated by the gray band. These
results are therefore not compatible. B3 and coeval experiments before
CHOOZ were all provided in units of kinetic positron energy, missing
the two annihilation~$\gamma$. We chose to use the visible energy
and therefore shifted up B3 spectra by 1.022~MeV for comparison.}
\end{figure}

Among all existing reactor antineutrino experiments, four of them
gives precise reactor spectra shape information. The Bugey~3 experiment
(B3)~\cite{B3} has until recently provided the finest reactor antineutrino
spectrum. The B3 measurement was in very good agreement with previous
predictions~\cite{FE,SC,AH}. The comparison is here updated to the
most recent predictions~\cite{MU,HU}. As indicated on Figure~\ref{fig:Ratios},
the net effect is an additional 1\%/MeV decrease through the full
energy range. This update is still compatible with prediction within
the 2\% linear spectral uncertainty envelop quoted in~\cite{B3}.
New measurements have been provided by three experiments: Double Chooz
(DC)~\cite{DC}, Daya Bay (DB)~\cite{DB} and RENO (RN)~\cite{RN}.
Their ratios to the state of the art prediction~\cite{MU,HU} are
depicted on Figure~\ref{fig:Ratios} and exhibit a significant deviation
from unity around 5~MeV. At first glance they clearly show a common
feature which is described as a bump in the 4~to 6~MeV region. Nevertheless,
to our knowledge, no quantitative comparison is available in the literature.

To gain quantitative insights on their compatibility, each spectrum
having different bin centers and widths, a direct $\chi^{2}$ comparison
is not possible. A bespoke statistical test was constructed to assess
if all the observed spectra could come from a common unique distribution.
This shared distribution was estimated thanks to a $\chi^{2}$ approach
using a Gaussian mixture model~\cite{HV}. Such an approach offers
the advantages of being reactor modeling independent and flexible
enough to accurately fit each single data set. To do so, a large number
($\mbox{K=42}$) of knots~$(x_{1},\ldots,x_{K})$ were used, evenly
distributed between 0\ and 10\ MeV. The spectral density, $f(x)=\frac{1}{h}\sum_{k=1}^{K}\,w_{k}\,\phi\left(\frac{x-x_{k}}{h}\right)$,
with weight parameters $w_{k}$, a bandwidth fixed to the inter-knot
distance $h=x_{k+1}-x_{k}$, $\phi$~being the standard normal probability
distribution function, was then integrated over each bin width to
properly model their content. On top of statistical uncertainties,
normalization and linear energy scale uncertainties were included
in covariance matrices. For DC a 1\% in normalization and scale was
used, for DB a 2.1\% and 1\%, for RN a 1.5\% and 1\% and for B3 a
4\% and 1.4\% as published by the collaborations~\cite{DC,DB,RN,B3}.
The $\chi^{2}$ was expanded with a quadratic term penalizing higher
local curvatures for smoothness: $\lambda\,\int f^{\prime\prime}(x)^{2}\,{\rm d}x$.
The amplitude of the positive parameter $\lambda$ controls the regularization
strength or equivalently the smoothness of the fitting function. It
was automatically determined from data through the generalized cross
validation method~\cite{GO}. This procedure is nearly equivalent
to minimizing the model predictive error. The optimization of the
global $\chi^{2}$ was iteratively performed until reaching convergence,
each step alternating between the $\chi^{2}$ function minimization
with respect to the $w_{k}$ parameters and the generalized cross
validation criterion minimization with respect to $\lambda$. The
model was fitted to all possible subset of experiments. While our
method was able to individually reproduce each spectrum accurately
adjusting differently the $w_{k}$ and $\lambda$ coefficients, it
prominently stressed out their inconsistency when combined. A parametric
bootstrap procedure was used to estimate the $\chi_{{\rm min}}^{2}$
distribution. For each experiment, the combined best fit model (gray
curve on Figure~\ref{fig:Ratios}) was used to draw $10^{4}$ Monte
Carlo simulations using their respective covariance matrix. Each of
these $10^{4}$ four~experiment data set was fitted with the same
global framework and the residual sum of squares was computed. Their
distribution was found to follow with a good accuracy a Gamma probability
density function with a shape parameter of $33\,(1\pm9\%)$ and a
scale parameter of $2.8\,(1\pm9\%)$. The associated p-value was estimated
using another $10^{6}$ Monte Carlo simulations, because of the uncertainties
on the Gamma PDF fitted parameters. The $99^{{\rm th}}$ percentile
of the p-value distribution was found below $10^{-5}$ while the median
was located around $10^{-8}$. As a prominent conclusion, the observed
spectra cannot come from a unique distribution with a high statistical
significance (more than $4.4\,\sigma$ at $99\%$ CL). This result
is mostly driven by the mismatch between the DB, RN and B3 spectra
since they have by far the highest statistics. Any combination of
two of these three experiments is also not consistent. Note that DB
and RN chose a different ad~hoc normalization. A quick integration
of their respective spectra seems to indicate a 2.9\% offset in their
respective normalization. Our aforementioned procedure outputs respectively
a median 3.4~$\sigma$ (4\,$\sigma$) incompatibility with (without)
this normalization adjustment. A normalization free fit (increasing
normalization uncertainties to infinity in both DB and RN covariance
matrices), yields a consistent result with a 3.4~$\sigma$ incompatibility
(p-value below $7\,10^{-4})$. Except for DC whose tests are compatible
with all other experiments within 1.6~$\sigma$, the lowest rejection
significance between pairs of experiments was found for DB and B3,
with a 2.5~$\sigma$ significance. A free norm fit still yielded
a 2.3~$\sigma$ significance rejection of compatibility hypothesis.

\subsection{On isotopic compatibility of reactor spectra}

On the antineutrino production side, all reactors are similar pressurized
water reactors. Seeking for differences among experiments, the core
isotopic composition is the main differing component which could play
a major role in the antineutrino spectral shape. For a complete fuel
burning (a 60-70\,GWd/t burnup), the reactor antineutrino rates are
known to roughly decrease by $-10\%$ and the spectra to tilt by approximately
$-4\%/{\rm MeV}\ensuremath{}$ between the fresh beginning and the
far end of nuclear fuel burning. This effect is reduced by a factor
3 to 4 taking into account reactor operations where fuel assemblies
are refreshed by thirds or quarters of the whole core. As indicated
in Table~\ref{tab:Fission-fractions-of}, the average isotopic compositions
over the data taking periods are very close from each other~\cite{DC,DB,RN,B3}.
On the one hand, DB and RN have isotopic compositions which differ
upmost by $1.7\%$ (mixture of 6~cores with comparable fuels), but
disagree on the distortion amplitude by 60\%. On the other hand DC
has the most burnt makeup among the four~experiments and displays
a distortion amplitude comparable to DB, while B3 reactor composition
is half-way between DC and DB but does not observe any distortion
in the spectrum. There is no simple coherent pattern explaining these
variations with isotopic compositions. In a more complex scenario
if the antineutrino spectra were extremely sensitive to fission fractions
from each reactor, the near/far relative measurement strategy for
the $\theta_{13}$ quest would have failed. With many cores and the
disparity of solid angle exposures among each detector, the near/far
ratios of DB and alike experiments would also be distorted from one
another to a large extent. 

As a conclusion, the differences in the published DC, DB, RN and B3
data-to-prediction ratios cannot be explained by core isotopic compositions
and therefore, must have another origin. Considering these distortions
are proportional to reactor powers~\cite{DC}, a remaining possibility
is to investigate at potential detector effects. While not being a
definitive argument, it seems more likely to distort an energy scale
and produce warped observed spectra to explain DC, DB and RN~\cite{DC,DB,RN}
than to wipe out a suspected spectral distortion with the exact required
compensation from the energy scale to explain the flatness of spectra
ratio in B3~\cite{B3}. In the next sections unassessed residual
nonlinearities in energy scales are studied as a potential scenario
to explain not only the difference between experiments but also the
mismatches in the data-to-prediction ratios.

\begin{table}
\begin{centering}
\begin{tabular}[b]{lcccc}
\hline 
\noalign{\vskip\doublerulesep}
 & $\phantom{}^{235}{\rm U}$ & $\phantom{}^{239}{\rm Pu}$ & $\phantom{}^{238}{\rm U}$ & $\phantom{}^{241}{\rm Pu}$\tabularnewline[\doublerulesep]
\hline 
\noalign{\vskip\doublerulesep}
\hline 
\noalign{\vskip\doublerulesep}
Bugey 3 & 53.8 & 32.8 & 7.8 & 5.6\tabularnewline[\doublerulesep]
\noalign{\vskip\doublerulesep}
\noalign{\vskip\doublerulesep}
Double Chooz & 49.6 & 35.1 & 8.7 & 6.6\tabularnewline[\doublerulesep]
\noalign{\vskip\doublerulesep}
\noalign{\vskip\doublerulesep}
Daya Bay & 58.6 & 28.8 & 7.6 & 5.0\tabularnewline[\doublerulesep]
\noalign{\vskip\doublerulesep}
\noalign{\vskip\doublerulesep}
RENO & 56.9 & 30.1 & 7.3 & 5.6\tabularnewline[\doublerulesep]
\hline 
\noalign{\vskip\doublerulesep}
\end{tabular}
\par\end{centering}
\caption{\label{tab:Fission-fractions-of}Fission fractions in~\% of Bugey~3,
Double Chooz, Daya Bay and R\textgreek{ENO} experiments.}
\end{table}

\section{Energy determination\label{sec:Energy-determination}}

The four~aforementioned experiments detect antineutrinos in liquid
scintillators through the inverse $\beta$~decay reaction: $\bar{\nu}_{e}+p\rightarrow e^{+}+n$.
In this process, most of the antineutrino energy is directly conveyed
to the positron with nuclear recoil effects totalizing an uppermost
0.1\%/MeV correction on positron/antineutrino energy. Antineutrino
spectra are therefore well acquired through counting and energy determination
of the detected positron. Such positrons deposit their energy in the
scintillator, with subsequent light emission through energy transfers
between solvent, primary and secondary fluors. The light yield scales
nonlinearly with the deposited energy (with a characteristic ${\cal O}(10\%)$
energy distortion below and above 1~MeV~\cite{DB}). Dedicated and
careful laboratory measurements are required for every liquid scintillator
to achieve an energy determination accuracy at ${\cal O}(1\%)$~\cite{AB,WA}.
The faint scintillation light is converted using photomultipliers
tube (PMTs) to measurable pC charge signals to estimate the energy
of the incident particle. Because of the data acquisition systems,
digitization process, detection threshold effects, width of time window
for charge acquisition, the typical nonlinearity in charge collection
amounts to ${\cal O}(10\%)$~\cite{DB}. These effects are peculiar
to each acquisition system~\cite{DC,DB} and special LED and $\gamma$
calibration runs are used to characterize the full electronic acquisition
chain.

To study the robustness and flexibility of the energy scale calibration,
the most stringent Daya Bay results along with their recent reactor
antineutrino spectrum deconvolution~\cite{DB}, corrected from $\theta_{13}$
oscillation effect, have been considered. The same studies could however
be directly applied to the other reactor experiments. The Daya Bay
experiment used dedicated $\gamma$~calibration sources as well as
$\phantom{}^{12}{\rm B}$ $\beta$~spectrum to estimate and constrain
the energy scale nonlinearities with an empirical model~\cite{DB}.
This model is a twofold factor, with a first term accounting for scintillation
physics, and a second one accounting for electronic nonlinearities.
Such an arbitrary function was designed to fit well the available
data points, as illustrated on Figures~\ref{fig:global-fit}~(a)
($\gamma$~sources) and~(b) ($\phantom{}^{12}{\rm B}$ $\beta$~spectrum).
The $\gamma$~data points correspond to the ratios between reconstructed
energies and those estimated from the best fitted empirical model~\cite{DB}.
A gray shaded area indicates the uncertainty envelope estimated from
the comparison of 5~empirical models describing the energy scale
nonlinearities~\cite{DB}. However, a fit to the relative energy
resolution was also provided: 
\begin{equation}
\frac{\sigma_{E}}{\left<E_{{\rm rec}}\right>}=\sqrt{a^{2}+\frac{b^{2}}{E}+\frac{c^{2}}{E^{2}}}\ ,
\end{equation}

with $a=0.015$, $b=0.087$ and $c=0.027$~\cite{CL} as the fitted
parameters. While $b$ is governed by photostatistics, $a$ and $c$
are related to systematic effects. Parameter $a$ is driven by spatial
and temporal variations throughout the detector while $c$ arises
from intrinsic PMT and electronic noise. They both assess that the
energy scale uncertainty is rather above 1.5\% than below 1\%~\cite{DB}.
Double Chooz obtained a comparable larger value $a=0.018$~\cite{DC}.
Taking into account these two systematic effect evaluations from $a$
and $c$ a more conservative systematic uncertainty corridor was also
displayed on Figure~\ref{fig:global-fit}~(a).

Because the scintillation quenching depends on the particle type,
$\gamma$ from radioactive sources and neutron captures, $e^{-}$
from $\phantom{}^{12}{\rm B}$ $\beta$~spectrum, $e^{+}$ from antineutrino
inverse $\beta$~decay reaction do not produce the same amount of
light for a given deposited energy. Tackling further associated nonlinearities
would require to have a complete Monte Carlo simulation with scintillator
content and \emph{Cherenkov} radiation modeling, which is not within
the scope of this article. As opposed to scintillation light emission,
electronic charge collection nonlinearity is identical whatever the
particle type is, and will be our focus in what follows.

\section{Residual nonlinearity matching}

\begin{figure*}[t]
\centering{}\includegraphics[scale=0.62]{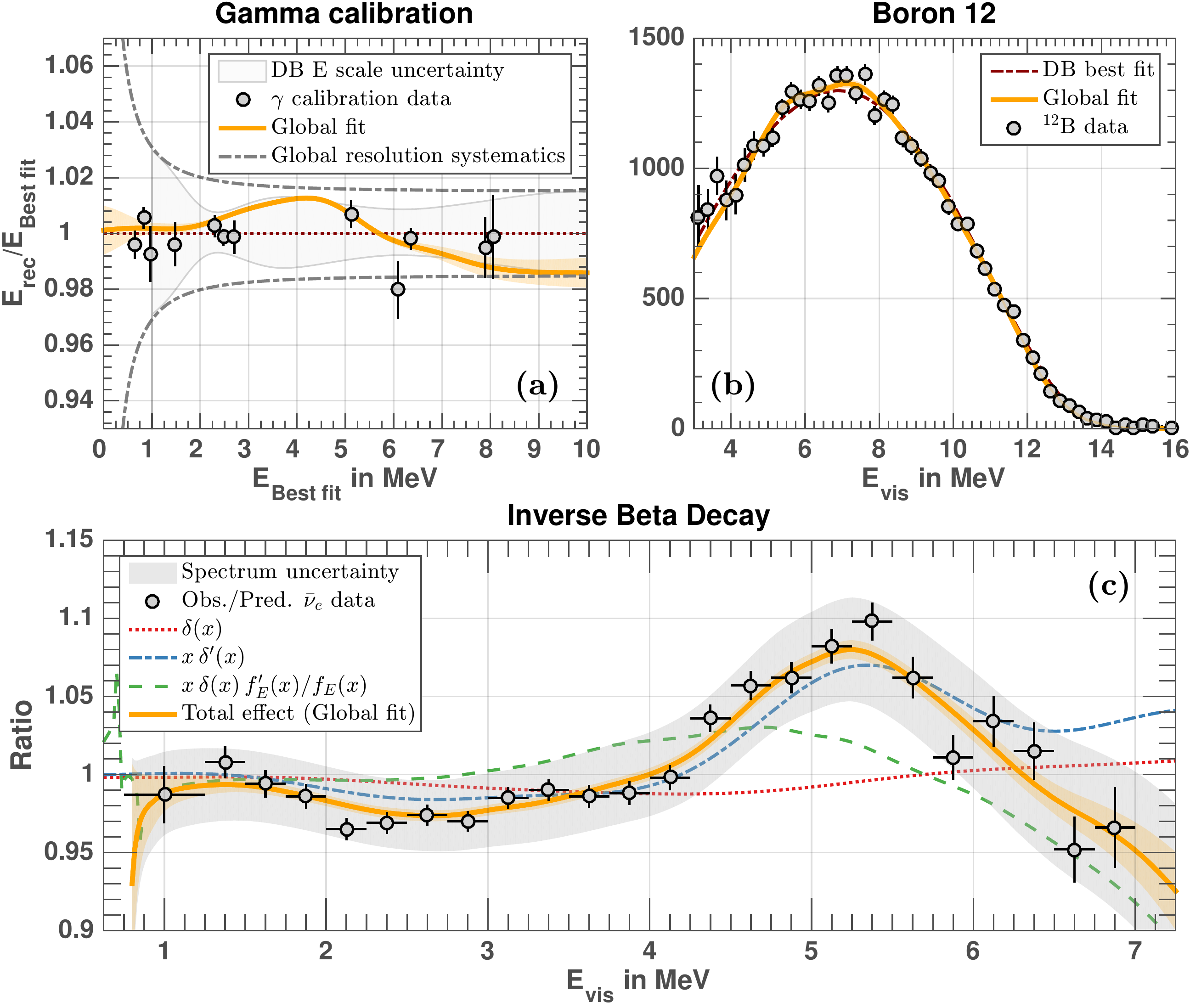}\caption{\label{fig:global-fit}Daya Bay's calibration and $\bar{\nu}_{e}$
spectrum ratio~\cite{DB}. (a) $\gamma$-calibration. They illustrate
the ratios between the reconstructed energies and the nominal best
fit from Daya Bay. (b) $^{12}{\rm B}$ $\beta$-decay spectrum is
used to demonstrate, to some extent, the validity of the calibration
process as in~\cite{DB}. (c) Inverse $\beta$-decay ($\bar{\nu}_{e}$)
spectra ratio between the observed spectrum ($\theta_{13}$ effect
removed) and the predicted one. The global best fit impact on (c)
is split in its 3~main components (upward shifted by +1 for display
purposes). The most prominent contribution is coming from the derivative
of the relative distortion $\delta$. A kink in energy scale around
4~MeV with an amplitude of nearly 1\% seen in (a) is sufficient to
explain a large fraction of the observed antineutrino spectral distortion
with respect to the prediction in (c) and still comply with (b). See
the text for further explanations of the involved mechanism.}
\end{figure*}

The aim of the present study is to assess if a small residual nonlinearity
(RNL) in the energy scale can both explain the observed reactor antineutrino
spectra and calibration data. For this purpose, an RNL function $\varphi$,
which adds more flexibility to the empirical model used by DB, was
introduced. As such, the probability density functions (PDF) of the
DB nominal reconstructed energies, $f_{E}$, and of the transformed
energies, $f_{\varphi(E)}$, are related through: $f_{E}(x)\,{\rm d}x=f_{\varphi(E)}\left(\varphi(x)\right)\,{\rm d}\varphi(x)$.
A more straightforward expression of the transformed energy PDF is:
\begin{equation}
f_{\varphi(E)}(x)=\frac{{\rm d}F_{E}\left(\varphi^{-1}(x)\right)}{{\rm d}x}\ ,\label{eq:transformed_PDF}
\end{equation}
where $F_{E}$ is the cumulative distribution function (CDF) of the
original energy variable. The RNL function may also be expressed through
the relative energy scale distortion~$\delta$:
\begin{equation}
\varphi(x)=x\,\left(1+\delta(x)\right)\ .
\end{equation}
The transformation relationship between the PDFs allows to simultaneously
determine $\varphi$ (or equivalently $\delta$) on $\gamma$~calibration,
$\phantom{}^{12}{\rm B}$ and $\bar{\nu}_{e}$ data. The function
$\varphi$ was estimated using a global $\chi^{2}$~fitting framework
on the 3~independent data sets and an additional regularization term
as in section~\ref{sec:Reactor-spectra-comparison}:
\begin{equation}
\chi^{2}=\chi_{\gamma}^{2}+\chi_{\nu}^{2}+\chi_{{\rm B}}^{2}+\chi_{{\rm R}}^{2}\ .\label{eq:global_chi2}
\end{equation}
The $\chi^{2}$ associated to $\gamma$~calibration data was defined
as
\begin{equation}
\chi_{\gamma}^{2}=\sum_{i=1}^{n^{(\gamma)}}\left(\frac{y_{i}^{(\gamma)}-\delta(x_{i}^{(\gamma)})}{\sigma_{i}^{(\gamma)}}\right)^{2}\ ,
\end{equation}
where $x_{i}^{(\gamma)},y_{i}^{(\gamma)}$ correspond to the $n^{(\gamma)}=12$
data points with associated uncertainties $\sigma_{i}^{(\gamma)}$,
as illustrated on Figure~\ref{fig:global-fit}~(a). The $\chi^{2}$
associated to $\bar{\nu}_{e}$~data was included as 
\begin{equation}
\chi_{\nu}^{2}=\sum_{i=1}^{n^{(\nu)}}\left(\frac{y_{i}^{(\nu)}-(1+\alpha_{\nu})\,N_{i}^{(\nu)}}{\sigma_{i}^{(\nu)}}\right)^{2}+\,\left(\frac{\alpha_{\nu}}{\sigma_{\nu}}\right)^{2}\ ,\label{eq:chi2_nu}
\end{equation}
where $y_{i}^{(\nu)}$ are the $n^{(\nu)}=24$ ratios of the observed
to predicted antineutrino spectra ($\theta_{13}$ effect already removed)
displayed on Figure~\ref{fig:global-fit}~(c), $\sigma_{i}^{(\nu)}$
the associated uncertainties, $\sigma_{\nu}$ an additional detector
normalization uncertainty of 2.1\%~\cite{DB} and $\alpha_{\nu}$
the corresponding nuisance parameter. Eventually, the $\chi^{2}$
associated to $\phantom{}^{12}{\rm B}$ spectrum was constructed as
follows:
\begin{equation}
\chi_{B}^{2}=\sum_{i=1}^{n^{(B)}}\left(\frac{y_{i}^{(B)}-\alpha_{B}N_{i}^{(B)}-\alpha_{N}N_{i}^{(N)}}{\sigma_{i}^{(B)}}\right)^{2}\ ,\label{eq:chi2_B}
\end{equation}
where $y_{i}^{(B)}$ are the $n^{(B)}=52$ experimental $\phantom{}^{12}{\rm B}$
spectrum bin values with uncertainties $\sigma_{i}^{(B)}$. The data
were freely fitted with both $\phantom{}^{12}{\rm B}$ and $\phantom{}^{12}{\rm N}$
components as in~\cite{DB}, as no further prior rate information
was available. Using Equation~(\ref{eq:transformed_PDF}), the $\bar{\nu}_{e}$
($N^{(\nu)}$), $\phantom{}^{12}{\rm B}$ ($N^{(B)}$) and $\phantom{}^{12}{\rm N}$
($N^{(N)}$) count rates are given by:
\begin{eqnarray}
N_{i}^{(\cdot)} & = & {\cal N}^{(\cdot)}\,\int_{x_{i}^{(\cdot)}-b_{i}^{(\cdot)}/2}^{x_{i}^{(\cdot)}+b_{i}^{(\cdot)}/2}f_{\varphi(E)}^{(\cdot)}\left(x\right){\rm \,d}x\nonumber \\
 & = & {\cal N}^{(\cdot)}\,F_{E}^{(\cdot)}\left(\varphi^{-1}\left(x_{i}^{(\cdot)}+\frac{b_{i}^{(\cdot)}}{2}\right)\right)\label{eq:N_PDF_fit}\\
 &  & -{\cal N}^{(\cdot)}\,F_{E}^{(\cdot)}\left(\varphi^{-1}\left(x_{i}^{(\cdot)}-\frac{b_{i}^{(\cdot)}}{2}\right)\right)\ ,\nonumber 
\end{eqnarray}
where $x_{i}^{(\cdot)}$ and $b_{i}^{(\cdot)}$ are the $i^{{\rm th}}$
bin center and width respectively, $F_{E}^{(\cdot)}$ is the associated
CDF in the original energy variable, $E$. ${\cal N}^{(\cdot)}$ is
a nominal normalization factor given in~\cite{DB}. $F_{E}^{(\cdot)}$
was determined from interpolation of Monte Carlo spectra from~\cite{DB}.
The target data set is indicated in the superscript parentheses~$^{(\cdot)}$
as in Equations~(\ref{eq:chi2_nu}) and~(\ref{eq:chi2_B}). The
$\bar{\nu}_{e}$ data-to-prediction ratio was then estimated using
the quotient between the above expression and the nominal DB prediction~\cite{DB}.
For small RNL ($\delta(x)\ll1$), $\varphi^{-1}$ can be expressed
as $\varphi^{-1}(x)\simeq x\,(1-\delta(x))$. A Gaussian mixture model~\cite{HV}
was chosen for modeling $\delta$ with $K$~evenly spaced knots,
$x_{k}$, between 0~and 16~MeV, a bandwidth, $h$, equal to the
inter-knot distance: $\delta(x)=\frac{1}{h}\sum_{k=1}^{K}\,w_{k}\,\phi\left(\frac{x-x_{k}}{h}\right)$,
where, $\phi$ is the standard normal probability distribution function.
The relative distortion just defined is rather flexible. To avoid
data overfitting, the last term in our global $\chi^{2}$ Definition~(\ref{eq:global_chi2})
is a quadratic regularization term penalizing higher local curvatures
in $\delta$ as described in section~\ref{sec:Reactor-spectra-comparison}:
$\chi_{R}^{2}=\lambda\int\delta^{\prime\prime}(x)^{2}\:{\rm d}x$.
The regularization level, $\lambda>0$, was self-determined from data
through the generalized cross validation method~\cite{GO}. The $\chi^{2}$
optimization was found to be still slightly sensitive to the number
and the position of the knots. To further prevent this, an extra regularization
procedure was used to complement the global $\chi_{R}^{2}$ penalization
term. While the bandwidth was kept fixed to the inter-knot distance,
the number of knots, $K$, was chosen with respect to the quality
of the standardized fit output residuals. It had to be large enough
to correctly model the distortions (more than $10$~knots) but small
enough to avoid overfitting (less than $30$~knots). An optimum of
$18$~knots was selected, corresponding to a standardized residual
distribution the closest to a standard normal distribution among our
investigations.

\section{Results}

Figure~\ref{fig:global-fit} presents the output of the global $\chi^{2}$~function~(\ref{eq:global_chi2})
optimization. The RNL best fit appears in orange thick solid line
with a shaded orange area indicating the $1\,\sigma$ uncertainty.
The best fit result gives $\chi_{\min}^{2}/{\rm ndof}=89.7/86$ (p-value
of $0.39$) with a standardized residual distribution following the
standard normal distribution. The fitted distortion correctly reproduces
the $\gamma$~calibrations, the $\phantom{}^{12}{\rm B}$ constraint
and the observed antineutrino spectrum warping. The best fit RNL agrees
within $1\,\sigma$ with the calibration data points. The $\phantom{}^{208}{\rm Tl}$
and $\phantom{}^{60}{\rm Co}$ around 2.5~MeV as well as the $\phantom{}^{16}{\rm O}^{\star}$
at 6~MeV are slightly off by less than $2.4\,\sigma$ from the best
fit. It is worth noting that the obtained RNL is still compatible
with the uncertainties originally quoted by the DB collaboration (gray
shaded area on Figure~\ref{fig:global-fit}(a)). Furthermore, the
frequently mentioned constraint of the $\phantom{}^{12}{\rm B}$~spectrum
is validated, with a fit comparable to the nominal DB one~\cite{DB}.
The ratio of observed to predicted antineutrino spectra is well reproduced
with such a fitted RNL. As shown by Figure~\ref{fig:global-fit}(c),
all data points fall within the gray shaded area representing the
reactor spectrum prediction uncertainty~\cite{HU,MU} from DB~\cite{DB}.
This study demonstrates that a 1\% energy scale distortion could result
in a 10\% $\bar{\nu}_{e}$~spectrum deformation.

This result can be explained by a closer look at the PDF ratio of
the $\bar{\nu}_{e}$ reconstructed energies using the best fit RNL
and the DB nominal one. To a good accuracy, such a ratio can be computed
from a Taylor expansion of Equation~(\ref{eq:transformed_PDF}):

\begin{equation}
\frac{f_{\varphi(E)}(x)}{f_{E}(x)}=1-x\,\delta(x)\,\frac{f_{E}^{\prime}(x)}{f_{E}(x)}-\delta(x)-x\,\delta^{\prime}(x)\,.\label{eq:PDF_ratio}
\end{equation}
The first term corresponds to the case of no residual distortion,
where $\varphi(x)=x$ and $\delta(x)=0$. The next term, $-x\,\delta(x)\,\frac{f_{E}^{\prime}(x)}{f_{E}(x)}$,
expresses the relative change of shape in the PDF due to $\delta$.
As illustrated by the dashed green curve on Figure~\ref{fig:global-fit}(c),
it gives a maximum positive deviation around 4.5~MeV where $\delta(x)<0$,
along with a major negative contribution after 6~MeV where $\delta(x)>0$.
The red dotted curve on Figure~\ref{fig:global-fit}(c) shows the
small contribution of the third term, $-\delta(x)$, to the ratio~(\ref{eq:PDF_ratio}).
Although the maximal 1.2\% distortion from $\delta(x)$ occurs around
4~MeV, the fourth term, $-x\delta^{\prime}(x)$, introduces a 6\%
positive deviation which is \emph{de facto} shifted to the 5~MeV
region. It dominates by far the PDF ratio as defined by Equation~(\ref{eq:PDF_ratio})
in the shoulder region. It is also worth noting that this term explains
as well a large fraction of the deficit in the 2\textendash 3~MeV
energy region (already corrected from $\theta_{13}$ effect~\cite{DB}).
The notable feature of this overall RNL is a $+1\%$ amplitude kink
around 4~MeV. No calibration point constrains the nonlinearity in
this region. The nearest one, from $\phantom{}^{12}{\rm C}$, shows
a slight excess around 5~MeV, which is perfectly reproduced by our
best fit nonlinearity. As shown by Figure~\ref{fig:global-fit}(b),
the Boron~12 spectrum is mostly unaffected by this additional RNL.

\section{Discussion}

The fit from previous section is the required RNL to best match the
observed antineutrino spectrum from the predicted one. It should be
understood as a residual artifact, or bias, after all calibration
works have already been performed. Our study shows that a small 1\%
mismatch in energy scale model in a localized energy range around
4~MeV, where no calibration cross-checks are available, can be responsible
of the large 10\% distortion in the observed to predicted antineutrino
spectra. This effect is thus a small one, within calibration uncertainties,
which might have been overlooked, and the consequences in reactor
spectra prediction are sizable. Bringing definitive statements about
absolute reactor spectra measurements requires extremely careful work
on energy scale to ensure its robustness. This indicates that shape
systematics in observed antineutrino spectra might have been underestimated.

Such a large distortion amplification process is rooted in two phenomena.
First, the energy scale uncertainties are relative. The higher the
energy, the higher the absolute uncertainty. A fixed relative energy
scale error hence induces a bigger impact at higher energies. A distortion
around 4~MeV in energy scale might thus be shifted above 5~MeV in
spectra ratio. Second, histograms are representative of a probability
density function. Under a change/distortion of the histogrammed variable,
they behave as a density function. Consequently an extra ``infinitesimal
volume conservation factor'' arises with the change of shape of the
density function. From these two characteristic effects, a localized
change of slope in energy scale around 4~MeV of less than 1\%/MeV
alters significantly the antineutrino counting distribution around
5\,MeV, as extensively demonstrated in previous section.

The origin of such an artifact in energy calibration models should
be investigated and tested. It could be of statistical nature, such
as a bias coming from an average of nonlinearly biased charges. All
the current energy reconstruction strategies in DC, DB, RN use the
total charge, received by all PMTs, as a proxy of the true deposited
energy. This work hypothesis, while fully legitimate when energy and
charge are proportional, starts to break down in scenarii accumulating
nonlinearities. Especially, if raw charge nonlinearities are not corrected
on each channel (a correction known as ``flat-fielding'' in digital
imaging), a complex position/energy nonlinear mapping arises and simple
calibration scheme, decoupling energy and position variables, might
be ineffective to simultaneously correct these dependences. As soon
as the raw charge distortion $\psi(Q)$ is not a linear mapping, the
average of the distorted charges is different from the distorted average
charge, whatever the distribution of $Q$ ($\langle\psi(Q)\rangle\neq\psi(\langle Q\rangle)$,
where $\langle.\rangle$ stands for the average over charge distribution).
In DC and DB, the electronics nonlinearity correction function is
a convex function of the reconstructed charge/energy. Therefore for
a given distortion $\psi$, the calibrations are overestimating the
true distortions since they use distorted charge averages to estimate
this distortion in E scale: $\psi(\langle Q\rangle)<\langle\psi(Q)\rangle$.
The fitted distortion from calibration data is thus overestimating,
by construction, the true raw charge distortion. The amount of overestimated
distortion depends on the peculiar charge distribution involved in
the average estimate, therefore on the energy and position inside
the detector. A part of this effect is canceled through the current
calibration procedures, especially below 3~MeV and by the fact that
in essence reconstructed energies are also to some extent average
of charges. However, because of the lack of calibration points around
4~MeV, a residual nonlinearity might still be present as we investigated
in this work.

This residual nonlinearity can be experimentally tested with dedicated
calibrations around 4~MeV and further works on raw charge corrections.
With the high statistics of the DB experiment it is also possible
to build up local antineutrino spectra inside the detectors. If charge
reconstructions are biased, the comparison of these local spectra
should exhibit enlarged variances, especially in the 4-5 MeV range.
The DB and DC fits to the energy resolution already support such a
point, with an energy scale systematics evaluated in DB around 1.5\%
and 1.8\% in DC, nearly two times bigger than the quoted precision
of the energy scale models. A systematic of such an amplitude is by
far sufficient to make observed and predicted antineutrino spectra
consistent with each other.

As a conclusion, current observed antineutrino spectra of DC, DB,
RN and B3 are not compatible with each other. This study suggests
that reactor $\bar{\nu}_{e}$ spectral distortions might have their
origin in detector calibration artifacts. Taking into account this
systematic effect, reactor spectra predictions become fully compatible
with all the observations. Our work is, to our knowledge, the single
one able to reconcile current generation of reactor antineutrino experiments
with older ones such as Bugey\,3, to recover consistency with BILL
experiments and their converted $\beta$~spectra. Because an energy
scale nonlinearity induces a migration of events across bins, it induces
mostly no impact on the reactor rate anomaly. Thus, if this interpretation
is correct, the current observed distortions in reactor antineutrino
spectra might have no relation with the 6\% reactor rate anomaly.
Regarding currently observed antineutrino spectra, as opposed to~\cite{HA},
spectra prediction uncertainties would remain unchanged, but observed
reactor spectra uncertainties of Double Chooz~\cite{DC}, Daya Bay~\cite{DB}
and RENO~\cite{RN} should be enlarged. As demonstrated in this article,
the 10\% spectral distortion around 5\,MeV is well within the 1~$\sigma$
calibration uncertainty of the energy scale. The uncertainties of
Bugey~3~\cite{B3} being already consistent with our fit (the slope
on spectra ratio might be induced by an energy scale bias within the
quoted uncertainties), no extra calibration error is justified for
this experiment. Decisive tests have to be planned. Until then a diligent
work on studying the energy response of detectors in the 3\textendash 5
MeV region is clearly indicated.\\
\\

\end{document}